\documentclass[preprint,amsmath,amssymb,aps,showkeys,showpacs]{revtex4}
\usepackage[english]{babel}
\usepackage{graphicx}
\usepackage{graphics}
\usepackage{amsmath}
\usepackage{dcolumn}
\usepackage{amssymb}
\usepackage{bm}

\begin{document}
\title{On the attractive inverse-square potential in the induced electric dipole system under the influence of the harmonic oscillator}
\author{K. Bakke}
\email{kbakke@fisica.ufpb.br}
\affiliation{Departamento de F\'isica, Universidade Federal da Para\'iba, Caixa Postal 5008, 58051-900, Jo\~ao Pessoa, PB, Brazil.}

\author{J. G. G. S. Ramos}
\email{jgabriel@fisica.ufpb.br}
\affiliation{Departamento de F\'isica, Universidade Federal da Para\'iba, Caixa Postal 5008, 58051-900, Jo\~ao Pessoa, PB, Brazil.}

\begin{abstract}

We obtain the analytical solutions to the Schr\"odinger equation for the attractive inverse-square potential in an induced electric dipole moment system under the influence of the harmonic oscillator. We show that bound states can exist when the electric field configuration brings a cut-off point that imposes a forbidden region for the neutral particle. Then, by dealing with $s$-waves, we obtain the energy eigenvalues in the strong electric field regime and for small values of the angular frequency of the harmonic oscillator. Further, we extend our discussion about the energy eigenvalues beyond the $s$-waves.

\end{abstract}


\maketitle

\section{Introduction}

In 1950, Case \cite{bessel1} brought a discussion about singular potentials in relativistic and non-relativistic quantum mechanics. In particular, Case dealt with the attractive inverse-square potential. Later, Landau and Lifshitz \cite{landau} extended this discussion and showed an unusual behaviour in the analysis of the bound state solutions, which they called as ``the fall of the particle to the center''. In their analysis, Landau and Lifshitz \cite{landau} considered a small region around the origin of radius $R$, which is known as the short-distance cut-off, and thus, showed that by taking the limit $R\rightarrow0$, the energy levels go to $-\infty$. Thereby, the short-distance cut-off guarantees that the spectrum of energy is finite \cite{bessel6,bessel9}. In recent decades, the attractive inverse-square potential has been studied through the Efimov effect \cite{squa5}, the generalized uncertainty principle \cite{squa6}, the electric dipole system in a (2+1)-dimensional conical spacetime \cite{bessel4} and the Aharonov-Bohm effect \cite{oliveira}. This singular potential has also been studied from the interaction of atoms with a ferromagnetic wire \cite{squa4} and from the interaction of the permanent magnetic dipole moment of a neutral particle with a magnetic field \cite{bf4}. Others studies have dealt with the attractive inverse-square potential from the interaction of the induced electric dipole moment of an atom with the electric field produced by a long charged wire \cite{squa3,squa2,bessel3,bf}.

In this work, we raise a discussion about the influence of the harmonic oscillator, in view of its ubiquitous role as the first contribution of additional interactions, on the attractive inverse-square potential that arises in the induced electric dipole moment system studied in Refs. \cite{squa3,squa2,bessel3,bf}. This system of a neutral particle with an induced electric dipole moment has also drawn attention due to the quantum effects associated with the interaction of the induced electric dipole moment with crossed magnetic and electric fields. Wei {\it et al} \cite{whw} obtained a geometric quantum phase and Furtado {\it et al} \cite{lin3} obtained Landau-type levels. It is worth citing other studies of the interaction of atoms with crossed electric and magnetic fields, such as, two-dimensional quantum rings \cite{dantas}, the quasi-Landau behaviour in atomic systems \cite{cross11,cross12}, systems of atoms and molecules \cite{cross2,cross3,cross10}, large electric dipole moments \cite{cross7}, doubly anharmonic-type oscillator \cite{bs} and in the rotating reference frame \cite{dantas1,ob5,ob6}. Thereby, in this work, we search for analytical solutions to the Schr\"odinger equation. We show that, even under the influence of the harmonic oscillator, the attractive inverse-square-type potential in the induced electric dipole system can yield bound state solutions when we take into account a field configuration that brings a cut-off point that imposes a forbidden region for the neutral particle.

The structure of this paper is: in section II, we introduce the attractive inverse-square potential that stems from the induced electric dipole moment of an atom with the electric field produced by a long non-conductor cylinder. Then, we analyse the influence of the harmonic oscillator on this attractive inverse-square potential. By dealing with $s$-waves, we show that analytical solutions to the Schr\"odinger equation can be obtained. We also obtain the eigenvalues of energy. Later, we extend our discussion about the energy levels beyond the $s$-waves; in section III, we present our conclusions.

\section{Attractive inverse-square potential in electric dipole system}

Let us consider the electric field produced by a uniform distribution of the electric charges inside an infinitely long non-conductor cylinder. This cylinder has a radius $R$. Then, at $r\,>\,R$, the electric field is given by 
\begin{eqnarray}
\vec{E}=\frac{\lambda}{r}\,\hat{r},
\label{1.1}
\end{eqnarray}
where $\lambda=\frac{\rho\,R^{2}}{2\epsilon_{0}}$ is a constant related to the uniform volume charge density $\rho$ inside the cylinder. Henceforth, we assume that the region $r\,<\,R$ is forbidden for the neutral particle. Hence, the radius of the long non-conductor cylinder plays the role of the cut-off point.

Then, when this electric field interacts with an atom, it induces an electric dipole moment $\vec{d}=\alpha\,\vec{E}$, where $\alpha$ is the atomic polarizability. The potential energy associated with the interaction of the electric field (\ref{1.1}) with the induced electric dipole moment of the neutral particle is given by \cite{bessel3,whw,lin3,dantas} 
\begin{eqnarray}
V\left(r\right)&=&-\vec{d}\cdot\vec{E}=-\,\alpha\,E^{2}\nonumber\\
&=&-\frac{\alpha\,\lambda^{2}}{r^{2}}.
\label{1.2}
\end{eqnarray} 
Therefore, in the region $r\geq R$, we have an attractive potential proportional to the inverse-square of the radial distance \cite{bessel1,landau,bessel6,squa2,bessel3}.  

In addition, let us assume that this electric dipole system is subject to the two-dimensional harmonic oscillator. This extension of the model is to include an interaction between the dipole and an external material environment. Our generalization can be modeled by an effective potential $V\left(r\right)$ due to the presence of matter. We focus on the main excitonic contribution of $V\left(r\right)$, i.e., on the harmonic term of the potential of interaction. Thereby, the Schr\"odinger equation becomes (with the units where $\hbar=1$ and $c=1$)
\begin{eqnarray}
i\frac{\partial\psi}{\partial t}=-\frac{1}{2m}\left[\frac{\partial^{2}}{\partial r^{2}}+\frac{1}{r}\frac{\partial}{\partial r}+\frac{1}{r^{2}}\frac{\partial^{2}}{\partial\varphi^{2}}+\frac{\partial^{2}}{\partial z^{2}}\right]\psi-\frac{\alpha\,\lambda^{2}}{r^{2}}\,\psi+\frac{1}{2}m\,\omega^{2}\,r^{2}\psi.
\label{1.3}
\end{eqnarray} 
The solution to the Schr\"odinger equation (\ref{1.3}) is given by $\psi\left(t,\,r,\,\varphi,\,z\right)=e^{-i\mathcal{E}t+i\ell\varphi+ip_{z}z}\,f\left(r\right)$, where $\ell=0,\pm1,\pm2,\ldots$ is the eigenvalue of $\hat{L}_{z}$, $p_{z}$ is a constant which is the eigenvalue of $\hat{p}_{z}$ and $f\left(r\right)$ is a function of the radial coordinate. Then, by substituting this solution into Eq. (\ref{1.3}), we obtain the radial equation:
\begin{eqnarray}
f''+\frac{1}{r}\,f'-\frac{\left[\ell^{2}-2m\,\alpha\,\lambda^{2}\right]}{r^{2}}f-m^{2}\omega^{2}\,r^{2}\,f+\left(2m\mathcal{E}-p_{z}^{2}\right)\,f=0.
\label{1.4}
\end{eqnarray}

From now on, we analyse the case where $\ell=0$, i.e., we shall work with $s$-waves. Let us take, without loss of generality, $p_{z}=0$ and define the parameters:
\begin{eqnarray}
\Lambda^{2}&=&2m\,\alpha\,\lambda^{2};\nonumber\\
\tau&=&2m\mathcal{E},
\label{1.5}
\end{eqnarray}
thus, we can rewrite the radial equation (\ref{1.4}) in the form:
\begin{eqnarray}
f''+\frac{1}{r}\,f'+\frac{\Lambda^{2}}{r^{2}}f-m^{2}\omega^{2}\,r^{2}\,f+\tau\,f=0.
\label{1.6}
\end{eqnarray}

Let us perform a change of variables given by: $x=m\omega\,r^{2}$. Thus, the radial equation (\ref{1.6}) becomes
\begin{eqnarray}
f''+\frac{1}{x}\,f'+\frac{\Lambda^{2}}{4x^{2}}f+\frac{\kappa}{x}\,f-\frac{1}{4}\,f=0,
\label{1.7}
\end{eqnarray}
where we have defined the parameter $\kappa$ as
\begin{eqnarray}
\kappa=\frac{\tau}{4m\omega}.
\label{1.7a}
\end{eqnarray}
The solution to Eq. (\ref{1.7}) is given in the form
\begin{eqnarray}
f\left(x\right)=\frac{c_{1}}{\sqrt{x}}\,M_{\kappa,\,i\frac{\Lambda}{2}}\left(x\right)+\frac{c_{2}}{\sqrt{x}}\,W_{\kappa,\,i\frac{\Lambda}{2}}\left(x\right),
\label{1.7b}
\end{eqnarray}
where $c_{1}$ and $c_{2}$ are constants. The functions $M_{\kappa,\,i\frac{\Lambda}{2}}\left(x\right)$ and $W_{\kappa,\,i\frac{\Lambda}{2}}\left(x\right)$ are the Whittaker functions of first and second kinds of imaginary order, respectively \cite{imag}.

Let us take a solution to Eq. (\ref{1.7}) regular at $x\rightarrow\infty$. Therefore, let us take $c_{1}=0$ and write Eq. (\ref{1.7b}) in the form:
\begin{eqnarray}
f\left(x\right)\propto\frac{1}{\sqrt{x}}\,W_{\kappa,\,i\frac{\Lambda}{2}}\left(x\right).
\label{1.8}
\end{eqnarray}

From now on, let us assume that the radius $R$ of the long nonconductor cylinder is very small. It corresponds to the short-distance cut-off $r=R$ used in Refs. \cite{bessel6,bessel9}. Since $R$ is very small, hence, we can assume that $x\ll1$. In search of bound state solutions, let us use the relation \cite{imag}:
\begin{eqnarray}
W_{\kappa,\,i\mu}\left(x\right)=\frac{\Gamma\left(2i\mu\right)}{\Gamma\left(\frac{1}{2}-\kappa+i\mu\right)}\,M_{\kappa,\,-i\mu}\left(x\right)+\frac{\Gamma\left(-2i\mu\right)}{\Gamma\left(\frac{1}{2}-\kappa-i\mu\right)}\,M_{\kappa,\,i\mu}\left(x\right),
\label{1.9}
\end{eqnarray}
where $\mu=\Lambda/2$. Note that, when $x\ll1$, we can write
\begin{eqnarray}
M_{\kappa,\,i\mu}\left(x\right)\rightarrow\,x^{i\mu+\frac{1}{2}}.
\label{1.10}
\end{eqnarray}
Thereby, we can rewrite Eq. (\ref{1.9}) as
\begin{eqnarray}
W_{\kappa,\,i\mu}\left(x\right)\approx\frac{\sqrt{x}\,\,\Gamma\left(2i\mu\right)}{\Gamma\left(\frac{1}{2}-\kappa+i\mu\right)}\,e^{-i\mu\,\ln\left(x\right)}+\frac{\sqrt{x}\,\,\Gamma\left(-2i\mu\right)}{\Gamma\left(\frac{1}{2}-\kappa-i\mu\right)}\,e^{i\mu\,\ln\left(x\right)}.
\label{1.11}
\end{eqnarray}
Next, let us simplify our discussion by defining the parameter:
\begin{eqnarray}
\beta=\frac{1}{2}-\kappa,
\label{1.12}
\end{eqnarray}
then, we rewrite Eq. (\ref{1.11}) in the form:
\begin{eqnarray}
W_{\kappa,\,i\mu}\left(x\right)\approx\frac{\sqrt{x}\,\,\Gamma\left(2i\mu\right)}{\Gamma\left(\beta+i\mu\right)}\,e^{-i\mu\,\ln\left(x\right)}+\frac{\sqrt{x}\,\,\Gamma\left(-2i\mu\right)}{\Gamma\left(\beta-i\mu\right)}\,e^{i\mu\,\ln\left(x\right)}.
\label{1.13}
\end{eqnarray}

Let us proceed our discussion by using the asymptotic formula \cite{abra,olver}:
\begin{eqnarray}
\Gamma\left(a\zeta+b\right)\rightarrow\sqrt{2\pi}\,e^{-a\zeta}\,\left(a\zeta\right)^{a\zeta+b-\frac{1}{2}},
\label{1.14}
\end{eqnarray}
which is valid for large values of $\left|\zeta\right|$, where $a>0$ and $b\in\mathbb{C}$ are both fixed. By taking the function $\Gamma\left(\beta+i\mu\right)$, we have that $\zeta=\beta$, $a=1$ and $b=i\mu$. Moreover, for the function $\Gamma\left(\beta-i\mu\right)$, we have that $\zeta=\beta$, $a=1$ and $b=-i\mu$. Observe that for small values of the angular frequency $\omega$, we have that $\kappa\gg1$ and, as a consequence, we have $\left|\beta\right|\gg1$. Thereby, with a fixed $\mu=\Lambda/2$, we obtain from Eq. (\ref{1.14}):
\begin{eqnarray}
\Gamma\left(\beta+i\mu\right)&\rightarrow& \frac{\sqrt{2\pi}\left(\beta\right)^{\beta-\frac{1}{2}}}{e^{\beta}}\,e^{i\mu\,\ln\left(\beta\right)};\nonumber\\
\Gamma\left(\beta-i\mu\right)&\rightarrow& \frac{\sqrt{2\pi}\left(\beta\right)^{\beta-\frac{1}{2}}}{e^{\beta}}\,e^{-i\mu\,\ln\left(\beta\right)}.
\label{1.15}
\end{eqnarray}
In addition, for a strong electric field, we can write
\begin{eqnarray}
\Gamma\left(2i\mu\right)&\rightarrow&\frac{\sqrt{2\pi}\,e^{-\pi\,\mu}}{\sqrt{2\mu}}\,e^{-i\left(2\mu-\mu\,\ln\left(4\mu^{2}\right)+\pi/4\right)};\nonumber\\
\Gamma\left(-2i\mu\right)&\rightarrow&\frac{\sqrt{2\pi}\,e^{-\pi\,\mu}}{\sqrt{2\mu}}\,e^{i\left(2\mu-\mu\,\ln\left(4\mu^{2}\right)+\pi/4\right)}.
\label{1.16}
\end{eqnarray}

Therefore, for $x\ll1$, after substituting Eqs. (\ref{1.15}) and (\ref{1.16}) into Eq. (\ref{1.13}) we find  
\begin{eqnarray}
W_{\kappa,\,i\mu}\left(x\right)&\approx& 2A\,\sqrt{x}\,\cos\left(2\mu+\mu\,\ln\left(\frac{\beta\,x}{4\mu^{2}}\right)+\frac{\pi}{4}\right),
\label{1.18}
\end{eqnarray}
where 
\begin{eqnarray}
A=\frac{e^{-\mu\pi+\beta}}{\sqrt{2\mu}\,\,\beta^{\beta-1/2}}.
\label{1.18a}
\end{eqnarray}
Note that Eq. (\ref{1.18}) agrees with Ref. \cite{imag}. By substituting Eq. (\ref{1.18}) into Eq. (\ref{1.8}), for small $x$, we have 
\begin{eqnarray}
f\left(x\right)&\approx& 2A\,\cos\left(2\mu+\mu\,\ln\left(\frac{\beta\,x}{4\mu^{2}}\right)+\frac{\pi}{4}\right).
\label{1.19}
\end{eqnarray}

We have assumed that the region $r<R$ is a forbidden region for the neutral particle, therefore, we impose that the wave function vanishes at $r=R$. Thereby, with $x_{0}=m\,\omega\,R^{2}$ we have the boundary condition:
\begin{eqnarray}
f\left(x_{0}\right)=0.
\label{1.20}
\end{eqnarray}
By substituting Eq. (\ref{1.19}) into Eq. (\ref{1.20}), we obtain
\begin{eqnarray}
f\left(x_{0}\right)=0\Rightarrow \cos\left(2\mu+\mu\,\ln\left(\frac{\beta\,x_{0}}{4\mu^{2}}\right)+\frac{\pi}{4}\right)=0.
\label{1.21}
\end{eqnarray}
Hence, with $\mu=\Lambda/2$, we obtain from Eq. (\ref{1.21}):
\begin{eqnarray}
x_{0}=\frac{\Lambda^{2}\,}{\beta}\,e^{\left(\frac{\pi}{2\Lambda}-2\right)}\,e^{2\pi\nu/\Lambda},
\label{1.22}
\end{eqnarray}
where $\nu=0,\pm1,\pm2,\pm3,\ldots$. Observe that the possible values of $\nu$ must satisfy the condition: $x_{0}\ll1$. This condition is satisfied when $\nu=-n$, where $n=1,2,3,\ldots$. Thereby, by using Eqs. (\ref{1.12}), (\ref{1.7a}) and (\ref{1.5}), the energy levels are given by 
\begin{eqnarray}
\mathcal{E}_{n}=\omega-\frac{4\,\alpha\,\lambda^{2}\,\,e^{\left(\frac{\pi}{2\sqrt{2m\,\alpha\,\lambda^{2}}}-2\right)}}{R^{2}}\,\,e^{-2n\pi/\sqrt{2m\,\alpha\,\lambda^{2}}}.
\label{1.23}
\end{eqnarray}

Hence, the spectrum of energy (\ref{1.23}) stems from the interaction of the induced electric dipole moment of the neutral particle with the electric field (\ref{1.1}) subject to the harmonic oscillator potential in the region $r>R$. The influence of the harmonic oscillator can yield $\mathcal{E}_{n}>0$ in contrast to Refs. \cite{bessel3,bf,bessel6,bessel9}, where $\mathcal{E}_{n}<0$. The presence of the harmonic term imposes the raising of the minimum of the potential and, consequently, the possibility of forming positive-defined non-evanescent modes. Observe that, as $R\rightarrow0$, then, $\mathcal{E}_{n}\rightarrow-\infty$. Therefore, no ground state exists when $R\rightarrow0$. This behaviour is analogous to the case of the fall of the particle to the center showed by Landau and Lifshitz \cite{landau} in the study of an electrically charged particle subject to the attractive inverse-square potential. Therefore, the spectrum of energy is finite due to the presence of the short-distance cut-off $r=R$ \cite{bessel6}.

In the case where $\ell\neq0$, bound states can be achieved if $2m\alpha\lambda^{2}\,>\,\ell^{2}$. Thereby, the parameter $\Lambda$ defined in Eq. (\ref{1.5}) must be rewritten as:
\begin{eqnarray}
\Lambda=\sqrt{2m\,\alpha\,\lambda^{2}}\rightarrow \Lambda=\sqrt{2m\,\alpha\,\lambda^{2}-\ell^{2}}.
\label{1.25}
\end{eqnarray}
Then, after following the steps from Eq. (\ref{1.5}) to Eq. (\ref{1.23}), we obtain the energy levels are determined in terms of $\Lambda$:
\begin{eqnarray}
\mathcal{E}_{n,\,\ell}=\omega-\frac{2\,\left[2m\,\alpha\,\lambda^{2}-\ell^{2}\right]\,e^{\left(\frac{\pi}{2\sqrt{2m\,\alpha\,\lambda^{2}-\ell^{2}}}-2\right)}}{m\,R^{2}}\,\,e^{-2n\pi/\sqrt{2m\,\alpha\,\lambda^{2}-\ell^{2}}}.
\label{1.26}
\end{eqnarray}
It is worth observing that Eq. (\ref{1.23}) is a particular case of Eq. (\ref{1.26}).

Finally, we should observe that the energy levels given in Eqs. (\ref{1.23}) and (\ref{1.26}) are no longer obtained if $\ell^{2}\,\geq\,2m\alpha\lambda^{2}$. In this case, the solution to Eq. (\ref{1.7}) is no longer given in terms of the Whittaker function of second kind of imaginary order.

\section{conclusions}

We have investigated the influence of the harmonic oscillator on the interaction of the induced electric dipole moment of a neutral particle with an electric field that gives rise to an attractive inverse-square potential. The electric field is produced by an uniform distribution of the electric charges inside an infinitely long non-conductor cylinder of radius $R$, where this radius has played the role of the short-distance cut-off \cite{bessel6,bessel9}. In Refs. \cite{bessel3,bf}, bound states associated with the attractive inverse-square potential (\ref{1.2}) have been obtained. In contrast to Refs. \cite{bessel3,bf}, by considering a strong electric field and small values of the angular frequency of the harmonic oscillator, we have seen that the presence of the harmonic oscillator potential modifies the spectrum of energy of an atom subject to the attractive inverse-square potential.  

An aspect of the spectrum of energy (\ref{1.23}) associated with the $s$-waves is: it decreases exponentially with $n$. Another aspect of the spectrum of energy is that $\mathcal{E}_{n}\rightarrow-\infty$ as $R\rightarrow0$, which is analogous to the fall of the particle to the center showed by Landau and Lifshitz \cite{landau}. 

We have also extended our discussion beyond the $s$-waves. We have shown that bound states are achieved when $\ell^{2}\,<\,2m\alpha\lambda^{2}$. Then, we have obtained the spectrum of energy.

An interesting perspective is to consider the presence of a linear topological defect inside the long non-conductor cylinder. Examples of these linear topological defects are disclinations and dislocations \cite{val,kat}. In recent decades, Aharonov-Bohm-type effects \cite{ab,pesk} have been studied in quantum systems associated with the presence of disclinations and dislocations in the elastic medium \cite{bf,bueno,amaro,amaro2,e3,fur6,bf2}. Therefore, with the presence of a disclination or a dislocation inside the non-conductor cylinder considered in the present work, we have the possibility of finding Aharonov-Bohm-type effects.

\acknowledgments

The authors would like to thank CNPq for financial support.

\end{document}